\overfullrule=0pt   \magnification=\magstep1  \input amssym.def
\font\huge=cmr10 scaled \magstep2  
\voffset.4in
\def\i{{\rm i}}  \def\si{\sigma}    
\def\eps{\epsilon} \def\la{\lambda}
\def\Z{{\Bbb Z}}  \def\Q{{\Bbb Q}}  \def\C{{\Bbb C}} \def\g{{\frak g}}

\font\smal=cmr7
\font\smit=cmmi7
\font\smcap=cmcsc10 
%

\centerline{{\huge Comments on nonunitary conformal field theories}}\medskip
\bigskip 

\centerline{{Terry Gannon}}
\medskip\centerline{{\it Department of Mathematical Sciences}}
\centerline{{\it University of Alberta}}
\centerline{{\it Edmonton, Canada, T6G 2G1}}\smallskip
\centerline{{tgannon@math.ualberta.ca}}  \bigskip\medskip


{\smal As is well-known, nonunitary RCFTs are distinguished from unitary ones in a number of
ways, two of which are that the
vacuum 0 doesn't have minimal conformal weight, and that the vacuum column of the
modular} {\smit S} {\smal matrix isn't positive. However
 there is another
primary field, call it} {\smit o}{\smal, which has minimal weight
and has positive} {\smit S} {\smal column.
We find that often there is a precise and useful relationship, which we call 
the Galois shuffle, between
 primary} {\smit o} {\smal and the vacuum; among other things this can
explain why (like the vacuum) its multiplicity in the full RCFT should be 1.
As examples we consider the minimal WSU(N) models. We conclude with
some comments on fractional level admissible representations of affine 
algebras. As an immediate consequence of our analysis, we get the classification
of an infinite family of nonunitary W$_3$ minimal models in the bulk.}\bigskip

\noindent{{\bf 1.\ Introduction}}\medskip

In any RCFT, each chiral sector ${\cal
H}_a$ (a module of the chiral algebra, i.e.\ of the
vacuum sector ${\cal H}_0$) has a Hermitian inner product
$\langle,\rangle_a$, obeying $\langle vx,y\rangle_a=\langle
x,\omega(v)y\rangle_a$ for $x,y\in {\cal H}_a$ and all $v$ in the
chiral algebra $V={\cal H}_0$, where $\omega$ is an anti-linear
anti-involution. In a {\it unitary} RCFT, this inner product is in
addition {\it positive-definite} (i.e.\ $\langle x,x\rangle>0$ when $x\ne 0$).  

 From the point of view of string theory (being a quantum field theory), 
the requirement of unitarity
seems natural\footnote{$^1$}{{\smal Perhaps though this belief is somewhat
naive. As pointed out to the author by Christoph Schweigert, in string theory
the matter CFT is coupled to the (super-)ghost CFT, and the latter is 
of course nonunitary. Physically, what is required is the unitarity of the
corresponding BRST cohomology, since the physical states live there. The
relationship between the (non)unitarity of the matter CFT and the unitarity
of that cohomology
space isn't obvious. For a careful review of the relation of CFT to string
theory, see e.g.\ [1], especially section 6.}}, and indeed much of the work on fusion
 rings and on what we'll call modular data has assumed it. But from the
 perspective of e.g.\ statistical models, at least on a torus, that 
requirement seems unnecessarily
 restrictive. Indeed, some of the better known RCFTs are nonunitary,
such as the Yang-Lee model ($c=-22/5$).  Presumably (but see [2]!), most RCFTs will
 be nonunitary --- for example the Virasoro minimal
model $(p,p')$, with central charge $c=1-6\,(p-p')^2/pp'$, is
 unitary iff $|p-p'|=1$.

The Wess-Zumino-Witten models [3], describing 
strings living on Lie group manifolds and corresponding to affine algebras at
positive integer levels (see e.g.\ [4]), will always be unitary, as will their GKO cosets
[5]. On the other hand, `admissible representations' [6] of affine algebras
at {\it fractional levels}  do not (yet) have a direct interpretation as an
RCFT, although their cosets have one as nonunitary RCFTs --- in fact this
is a powerful way to construct nonunitary theories. For example, the Virasoro
minimal $(p,p')$ model corresponds to the coset $(\widehat{{\rm su}}(2)_m\oplus
\widehat{{\rm su}}(2)_1)/\widehat{{\rm su}}(2)_{m+1}$, where $m={p'\over
p-p'}-2\in{\Bbb Q}$.

In this paper we compare nonunitary and unitary RCFTs.
We isolate in \S3 what seems to be a key uncertainty in {\it
nonunitary} theories:
the multiplicity in the RCFT (what we will denote below mult$_n(o)=M_{oo}$) 
of the primary $o$ with
minimal conformal weight. We learn that in any RCFT it will be
greater than 0, but the question of its precise value turns out to be
important. We conjecture in \S4 that this multiplicity should equal 1; this would
imply, for {\it nonunitary RCFT},  irreducibility of {\smcap nim}-reps and finiteness results for modular
invariants and {\smcap nim}-reps, among other things. This multiplicity issue did
not arise in the Virasoro minimal model classification [7], because
as we will see in \S4, modular invariance there forces this multiplicity to be 1. 

We find that in many theories,  the primary $o$ is related to the vacuum
0 in a certain simple way --- we'll say such theories possess the {\it Galois
Shuffle} (or {\it GS}) property. In particular this holds for the
 nonunitary $W_N={WSU}(N)$ and $WSO(2N)$ minimal
models, at arbitrary rank $N$ and arbitrary values $p,p'$ (these
include of course the Virasoro minimal models). 
Related comments are made for the affine algebras at fractional
level. Because of their role in the quantum
Drinfeld-Sokolov reduction [8], this suggests that the GS property may be
fairly typical among RCFTs.

Nevertheless, we know that not all all RCFTs possess this property.
When the GS property holds for a given RCFT, there are many nontrivial consequences,
as we will see in \S\S4 and 6. One of these will give us for free a
nonunitary RCFT classification. Let us briefly describe another consequence.

First, recall that there are several unitary matrices $\widetilde{S}$
which diagonalise the fusion rules, i.e.\ for which the fusion coefficients
$N_{ab}^c$ can be formally recovered from Verlinde's formula when the modular matrix
$S$ is replaced with $\widetilde{S}$. In particular take the modular
matrix $S$ and permute the columns, and multiply each column by an arbitrary
phase. But typically the symmetry condition $S=S^t$ will be lost. However,
there are two generic ways to find {\it symmetric} unitary matrices $\widetilde{S}$
which diagonalise the fusions: simple-currents and the Galois
symmetry. We will discuss these generic constructions  next section.

Now, the GS proprty implies that the fusion ring of the nonunitary theory can be
diagonalised by a symmetric $S$-matrix $\widehat{S}$ which obeys all the usual properties
of a {\it unitary} theory (e.g.\ the 0th  column is strictly positive), and
that $\widehat{S}$ will be obtained from the `sick' modular
$S$ matrix by a generic construction (a `Galois shuffle').
(By `sick' here we only mean that the vacuum column of $S$ is not positive.) 
We call $\widehat{S}$ the {\it unitarisation} of the nonunitary theory.
In other words, {\it the fusion rings of nonunitary and unitary theories are
indistinguishable,} at least when the nonunitary theory obeys the GS property.

To give a simple explicit 
example,  the fusions of the nonunitary Yang-Lee model are diagonalised
of course by its `sick' modular $S$ matrix
$$S={2\over \sqrt{5}}\left(\matrix{-\sin(2\pi/5)&\sin(\pi/5)\cr \sin(\pi/5)
&\sin(2\pi/5)}\right)\ .$$
But the Yang-Lee fusions are also diagonalised by the unitarisation
$$\widehat{S}={2\over \sqrt{5}}\left(\matrix{\sin(\pi/5)&\sin(2\pi/5)\cr
\sin(2\pi/5)&-\sin(\pi/5)}\right)\ ,$$
which is the modular $S$ matrix of the (unitary) WZW model for the affine
algebras $\widehat{G}_2$ and $\widehat{F}_4$
at level 1. However, as we will discuss below,
the unitarisation will not in general correspond to a WZW model. Incidentally,
the tensor product of the $\widehat{G_2}{}_{,1}$ WZW model with the Yang-Lee
model is an example of a theory which doesn't obey GS.

What does this unitarisation mean at the level of the RCFT? Indirectly, we
can see the corresponding unitary theory in e.g.\ the fusion ring and the
list of cylindrical and toroidal partition functions. Typically these partition
functions will be in one-to-one correspondence with those of the nonunitary 
theory, and we will use this
correspondence below to obtain the classification of a new infinite family
of nonunitary theories. But can we see the unitarisation
{\it directly} inside the nonunitary theory, perhaps at the level of the
chiral algebras? This is still unclear.

\bigskip\noindent{{\bf 2.\ Background material}}\medskip

The characters $\chi_a(\tau)={\rm Tr}_{{\cal H}_a}q^{L_0-c/24}$ of an RCFT carry a representation of
SL$_2(\Z)$:
$$\eqalignno{\chi_a(-1/\tau)=&\,\sum_bS_{ab}\,\chi_b(\tau)\ ,&(1a)\cr 
\chi_a(\tau+1)=&\,\sum_bT_{ab}\,\chi_b(\tau)\ .&(1b)} $$
The subscripts $a,b$ here label the (finitely many) chiral primary fields $\phi$.
The `modular matrices' $S$ and $T$ are unitary and symmetric, and $T$ is
diagonal with entries exp$[2\pi\i\,(h_a-{c\over 24})]$ where $h_a$ are the
conformal weights and $c$ is the central charge. The matrix $T$ has finite order,
i.e.\  all numbers $h_a$ and $c$ are rational. The matrix
$C=S^2$ is the order-2 (or order-1) permutation matrix called charge-conjugation.
Then $(ST)^3=(TS)^3=C$.
Moreover, $$S_{ab}^*=S_{Ca,b}=S_{a,Cb}\ .\eqno(1c)$$
These properties should hold for any RCFT [9]. 

The fusion coefficients $N_{ab}^c$ can be expressed using
Verlinde's formula [10]:
$$N_{ab}^c=\sum_d{S_{ad}\,S_{bd}\,S^*_{cd}\over S_{0d}}\ ,\eqno(2)$$
where `0' denotes the vacuum. We have $h_0=0$ and $C0=0$. It is convenient to define the {\it
fusion matrices} $N_a$ by $(N_a)_{bc}=N_{ab}^c$. Then Verlinde's
formula says that $S$ simultaneously diagonalises all matrices $N_a$, with
eigenvalues $S_{ab}/S_{0b}$.

In the case of nonunitary RCFTs (as discussed next section), this
so-called `vacuum' 0 does not possess all the usual properties expected of a
physical vacuum, and so perhaps
a better name for it would be `identity'.

\medskip \noindent {\bf Definition 1.} By {\it modular data} we mean here
unitary matrices $S,T$ such that $T$ is diagonal and of finite order,
$S$ is symmetric, $(ST)^3=S^2=:C$ is an order-2 (or order-1) permutation matrix,
$C0=0$, and
the fusion coefficients $N_{ab}^c$ given by (2) are all nonnegative
integers $\Z_+:=\{0,1,2,\ldots\}$.\medskip

See e.g.\ [11,12] for some of the basic properties of modular data,
although those papers primarily specialise to what we call below {\it
unitary} modular data (see Def.2 next section). There are many additional
properties which the modular data of a healthy RCFT should obey. For one
important example [13], for any primaries $a,b\in\Phi$, the numbers defined by
$$Z(a,b):=T_{00}^{{1\over 2}}T_{bb}^{*{1\over 2}}\sum_{c,d\in\Phi}
N_{cd}^aS^*_{bc}S_{0d}T^2_{dd}T^{-2}_{cc}$$
must always be integers satisfying
$$\eqalignno{|Z(a,b)|\le &\,N_{aa}^b\ ,&(3a)\cr Z(a,b)\equiv&\,N_{aa}^b\ ({\rm
mod}\ 2)\ .&(3b)}$$
In particular, $Z(a,0)\in\{\pm 1,0\}$ is the Frobenius-Schur indicator.
Also, [14] gives formulas for the traces of $S$, $ST$ and $(ST)^2$, among
other things. These identities and conditions won't play any special role in
this paper, for reasons we'll give later, and so are ignored in Def.1. We
turn next to an additional condition which will be used below. 

 Form the vector $\vec{\chi}(\i):=(\chi_a(\i))$ (each character
 converges throughout the upper half-plane, so these character values are
 defined). Since each character
$$\chi_a(\tau)=q^{h_a-c/24}\sum_{n=0}A_{a,n}q^n$$
has nonnegative coefficients $A_{a,n}$, each component $\chi_a(\i)$ of 
$\vec{\chi}(\i)$ will be strictly positive. Because $\tau=\i$ is a
fixed-point of the transformation $\tau\mapsto -1/\tau$, $\vec{\chi}(\i)$ will be an
eigenvector of $S$ with eigenvalue 1:
$$\sum_bS_{ab}\,\chi_b(\i)=\chi_a(\i)\ .$$

\noindent{\bf Fact 1.} $S$ has a strictly positive eigenvector,
corresponding to eigenvalue 1.\medskip

The one-loop torus partition function [15] 
$${\cal Z}(\tau)=\sum_{a,b}M_{ab}\,\chi_a(\tau)\,\chi_b(\tau)^*$$
is modular invariant: ${\cal Z}(-1/\tau)={\cal Z}(\tau)={\cal
Z}(\tau+1)$ etc, so the coefficient matrix $M$ lies in the commutant of the
SL$_2(\Z)$ representation: $SM=MS,TM=MT$. We also know that the
coefficients $M_{ab}$ are multiplicities, and so lie in $\Z_+$. Also,
the multiplicity of the vacuum must be 1: $M_{00}=1$.

A vacuum-to-vacuum amplitude of fundamental importance in
boundary CFT, is the cylindrical partition function (see for instance
[16,17,18,19] and references therein)
$${\cal Z}_{\alpha\beta}(t)=\sum_a n_{a\alpha}^\beta\,\chi_a(\i t)\ .$$
The coefficients $n_{a\alpha}^\beta$ will be nonnegative integers. 
Define matrices $n_a$ by
$(n_a)_{\alpha\beta}=n_{a\alpha}^\beta$. These matrices are required
to form a  representation, called a {\it {\smcap nim}-rep},  of the fusion ring: $n_an_b=\sum_c
N_{ab}^c n_c$. We can simultaneously diagonalise the $n_a$: equation (2) is
replaced with
$$n_{a\alpha}^\beta =\sum_d {\psi_{\alpha d}\,S_{ad}\,\psi_{\beta
d}^*\over S_{0d}}\ ,\eqno(4)$$
where each $d\in\Phi$ appears in (4) with a certain multiplicity mult$_n(d)\ge
0$ independent of $a,\alpha,\beta$.

There is a compatibility condition between the toroidal and cylindrical partition
functions: the multiplicity mult$_n(d)\in \Z_+$ of each primary $d$ in (4) must
equal the diagonal entry $M_{dd}$ of the   corresponding modular invariant $M$.
For instance, the diagonal modular invariant $M=I$ corresponds to the
fusion matrix choice $n_a=N_a$.

Two subtleties which we will avoid this paper are the gluing
automorphism of [20], and the fact that the characters $\chi_a(\tau)$
are usually not linearly independent. These play no special role
in our discussion, and the interested reader can consult the relevant literature.
To fix notation, note that we have chosen $S$ to correspond to
the matrix $\left(\matrix{0&-1\cr 1&0}\right)$ rather than its inverse,
which also appears in the literature.

Next, let's review the lesser known 
Galois action of RCFTs [21]. The entries of $S$
will lie in some cyclotomic extension of $\Q$ --- that just means each entry
$S_{ab}$ can be written as a polynomial $p_{ab}(x)$, with rational coefficients,
evaluated at some root of unity $x=\xi_L:=\exp[2\pi\i/L]$. In fact, $L$ can be taken
to be the order of the matrix $T$ [22,23]. By the `Galois
group' of this cyclotomic extension, we mean the multiplicative group of
integers mod $L$
which are coprime to $L$ --- we write this $\Z_L^\times$. For example,
$\Z_{12}^{\times}=\{1,5,7,11\}$. Any such
$\ell\in\Z_L^\times$ acts on the number $p_{ab}(\xi_L)$ by sending it to
$p_{ab}(\xi_L^\ell)$ --- we denote  this `Galois automorphism' by $\si_\ell$.
For instance: $\si_\ell$ fixes all rationals; $\sigma_\ell\,\cos(2\pi{a\over L})=\cos(2\pi{\ell a\over L})$;
$\si_\ell \,\i=\pm \i$ for $\ell\equiv \pm 1$ (mod 4), respectively;
and $\si_\ell\sqrt{2}=\pm\sqrt{2}$ depending on whether or not $\ell\equiv \pm 1$
(mod 8). What is so special about the $\sigma_\ell$ is that they are 
symmetries of the cyclotomic numbers: $\sigma_\ell(u+v)=\sigma_\ell(u)+\sigma_\ell(v)$ and
$\sigma_\ell(uv)=\sigma_\ell(u)\,\sigma_\ell(v)$, for any numbers $u,v$
in the $L$th cyclotomic extension $\Q[\xi_L]$ of $\Q$.

The point [21] is that to any $\ell\in\Z_L^\times$, there is a permutation
$a\mapsto \si_\ell a$ of the primary fields, and a choice of signs
$\epsilon_\ell(a)=\pm 1$, such that 
$$\si_\ell(S_{ab})=\epsilon_\ell(a)\,S_{\si_\ell
a,b}=\epsilon_\ell(b)\,S_{a,\si_\ell b}\ .\eqno(5a)$$

This Galois action should be regarded as a generalisation of charge-conjugation.
In particular, $\ell=-1$ corresponds to the familiar Galois automorphism $\si_{-1}=*$
(complex conjugation), with signs $\eps_{-1}(a)=+1$ and permutation
given by the charge-conjugation $\si_{-1}a=C{a}$. It was proved in  [22] that
 $$h_{\si_\ell
a}-c/24\equiv \ell^2\, (h_a-c/24)\qquad({\rm mod}\ 1)\ ,\eqno(5b)$$ 
where  we can take $\ell$ to be coprime to the
order $L$ of the matrix $T$ (in fact it suffices here to take  $\ell$ in (5b) 
coprime to the
denominator of the rational number $h_a-c/24$). The partition functions $M$ and $n$
respect this Galois action [21,24]:
$$\eqalignno{M_{ab}=&\,M_{\si a,\si b}\ ,&(5c)\cr {\rm mult}_n(a)=&\,{\rm
mult}_n(\si a)\ .&(5d)}$$

As mentioned in the Introduction, there are infinitely many different unitary
matrices $\widetilde{S}$ which diagonalise the fusion coefficients. We conclude this
section by mentioning a generic way to construct {\it symmetric} unitary matrices
$\widetilde{S}$. Choose any simple-current $J$ and any Galois automorphism
$\sigma_\ell$, where the integer $\ell$ is coprime to the order $L$ of $T$.
Define the following matrices $$\eqalignno{
\widetilde{S}_{ab}:=\si_\ell(S_{Ja,Jb})=&\,\exp[2\pi\i\,\ell\,(Q_J(b)+Q_J(J))]
\,\epsilon_\ell(b)\,S_{a,\sigma_\ell J b}\ ,&(6a)\cr
\widetilde{T}_{ab}:=&\,\delta_{ab}\,(T_{Ja,Jb})^\ell\ .&(6b)}$$
This implies $\widetilde{S}=\si_\ell(PSP^t)$
and $\widetilde{T}=\si_\ell(PTP^t)$, where $P$ is the
(orthogonal) permutation matrix $P_{ab}=\delta_{b,Ja}$, 
and these define a representation   of the modular group SL$_2(\Z)$: 
$(\widetilde{S}\widetilde{T})^3=\widetilde{S}^2=C$. This new representation
will typically be inequivalent to that of the original $S,T$. Note that
$\widetilde{S}$ is manifestly symmetric. {That $\widetilde{S}$
diagonalises the fusions can be verified either by direct calculation (using the
fact that the fusion coefficients are rational numbers), or by noting from
the right-side of (6a) that the $b$th column  of $S$ has been permuted by
$\sigma_\ell\circ J$, and multiplied by the phase
$\exp[2\pi\i\ell\,(Q_J(b)+Q_J(J))]\,\epsilon_\ell(b)$. 
Any such $\widehat{S},\widehat{T}$ will automatically obey the constraints
(3) and identities in [14], provided $S,T$ do, and provided we choose the
squareroot $\widehat{T}^{{1\over 2}}:=PT^{{\ell\over 2}}P^t$ (this is why
nothing will be gained in this paper by manifestly imposing those conditions
on modular data).
Of course an additional (trivial and well-known) construction is to tensor $\widetilde{S},\widetilde{T}$
by any 1-dimensional representation of SL$_2(\Z)$, i.e.\ to choose any (not necessarily
primitive) 6th
root $t$ of unity and to replace $\widetilde{S}\mapsto t^{-3}\widetilde{S}$,
$\widetilde{T}\mapsto t\widetilde{T}$ ($t$ should be a 
6th root of unity, in order to preserve  the positivity of $C$).

At least sometimes, the matrices $\widetilde{S},\widetilde{T}$ in (6) can
be realised by `characters'. Let $L$ be the order of $T$, and suppose $\ell$ is a perfect
square $m^2$ (mod $L$). For instance if $L$ or $L/2$ is a power of any odd prime,
then
half of the $\ell$'s in $\Z^\times_L$ would be  perfect squares mod $L$.
Let $m'$ be a multiplicative inverse (mod $L$) of $m$, and choose any matrix
$A\in {\rm SL}_2(\Z)$ obeying
$$A=\left(\matrix{a&b\cr c&d}\right)\equiv \left(\matrix{m&0\cr 0&m'}\right)
\qquad({\rm mod}\ L)\ .$$
Such an $A$ will always exist and, with respect to the SL$_2(\Z)$ representation
defined by (1), will correspond to the matrix $G_m$ defined by [23,22]
$$(G_m)_{ab}= (ST^{m'}ST^mST^{m'}C)_{ab}=\epsilon_{m}(a)\,\delta_{b,\sigma_\ell(a)}\ .$$
Define the functions
$$\widetilde{\chi}_{a}(\tau):=\chi_{J^{m'}a}({a\tau+b\over c\tau+d})=
\epsilon_m(a)\,\chi_{J\sigma_ma}(\tau)\ .\eqno(6c)$$
Then  these `characters' realise our matrices $\widetilde{S},\widetilde{T}$
in the sense of (1):
$$\eqalign{\widetilde{\chi}_a(-1/\tau)=&\,\sum_b\widetilde{S}_{ab}\,\widetilde{\chi}_b(\tau)\ ,\cr 
\widetilde{\chi}_a(\tau+1)=&\,\sum_b\widetilde{T}_{ab}\,\widetilde{\chi}_b(\tau)\ .}$$
Note that the coefficients of these `characters' $\widetilde{\chi}$
 are integers but may not be positive --- in fact, $\widetilde{S}$
may not satisfy {Fact 1} even if this `character' interpretation exists.

More generally however (e.g.\ for the Yang-Lee model of \S1 or the $c=-25/7$ minimal model
studied in \S6), the modified modular matrices $\widehat{S},\widehat{T}$
won't be realised by merely permuting the characters (i.e.\ reordering the
primaries), and their direct meaning in terms of the nonunitary theory
is still unclear. 

To our knowledge, equations (6) and the related remarks are new (but for
some thoughts in this direction see \S5 of [25]).

\bigskip\noindent{{\bf 3. Unitary vrs nonunitary RCFTs}}\medskip

In this section we describe the distinctions between the more familiar
unitary RCFTs, and the more complicated and common
nonunitary RCFTs.

The assumption of unitarity certainly does yield some simplifications. For
instance\footnote{$^2$}{{\smal I thank Matthias Gaberdiel for helpful
discussions on this point.}}, as is well known, the conformal weight $h=1$ fields
in the vacuum sector will always
form a (complex) Lie algebra ${\frak g}$, and the inner product $\langle,
\rangle_0$ will define an invariant symmetric bilinear form for ${\frak g}$.
When the RCFT is {\it unitary}, the positive-definiteness of that invariant
form implies that ${\frak g}$ will be
{\it reductive} [26,27] --- i.e.\ a direct sum   of an abelian Lie algebra
$\C^m$ with a number of simple Lie algebras. Otherwise, when the theory is
{\it nonunitary}, that Lie algebra belongs to a much wider class called
{\it self-dual}. Any such  Lie algebra can be constructed from a reductive
Lie algebra by a sequence of `double extensions' [28] (see also [29]). 
Self-dual Lie algebras have appeared
in the physics literature: for instance, the $2+1$-dimensional Poincar\'e
group ISO(2,1) used by Witten
[30] to relate Chern-Simons with 2+1-dimensional gravity has a self-dual Lie
algebra; in [31,32,33] and references therein, families of self-dual Lie
algebras have been explicitly related to CFT. Self-dual Lie algebras are
precisely those for which the Sugawara construction works.

We are more interested here in the differences at the level of modular
data, and at the level of the toroidal $M$ and cylindrical $n_a$ partition
functions. From this perspective, the key observation is that, in a
unitary theory, the vacuum 0 is the unique primary with minimal
conformal  weight $h$. Hence for $\tau=\epsilon \i$ ($\epsilon\approx 0$)
$$0<\chi_a(\tau)=\sum_bS_{ab}\,\chi_b(-1/\tau)\approx S_{a0}\,\chi_0(-1/\tau)\eqno(7)$$
and thus $S_{a0}=S_{0a}>0$. 

\medskip\noindent{{\bf Definition 2.}} By {\it unitary modular data}, we
mean modular data obeying the additional requirement that $S_{a0}>0$
for all $a\in \Phi$.\medskip

A word of warning should be made: the modular data of any unitary RCFT will
be unitary modular data. However, the converse may not hold: given some
unitary modular data, there may or may not be a unitary RCFT which realises
it --- e.g.\ Fact 1 or constraint (3) may be violated.

In any modular data, $S_{a0}$ is a nonzero real number (this follows because
$S^*_{a0}=S_{a,C0}=S_{a0}$).
What replaces the inequality $S_{a0}>0$  in a nonunitary theory?

First note that there must be one and only one chiral primary, call it $o$,  with 
$$S_{ao}=S_{oa}>0\qquad\forall a\ .\eqno(8)$$
To see this, number all the primaries
$a_1=0,a_2,\ldots,a_n$, and take $m\in {\Bbb R}$ to be large enough that the sum
$N:=\sum_{i=1}^nm^i\,N_{a_i}$ of fusion matrices has distinct eigenvalues
$\sum_i m^i\,S_{a_i b}/S_{0b}$ (in fact all these eigenvalues will be distinct,
for all but finitely many values of $m$). Then for some phase $\varphi$ and some
primary $o\in\Phi$, the column $\varphi S_{\updownarrow o}$ will be
the unique (strictly positive) Perron-Frobenius eigenvector [34] of the
strictly positive matrix $N$. But {Fact 1} implies that the phase $\varphi$ equals
$+ 1$ (if we don't use {Fact 1}, i.e.\ for arbitrary modular data not necessarily
corresponding to an RCFT, we only get that $\varphi=\pm 1$). By unitarity of $S$, there
can be only one chiral primary obeying (8).

\medskip\noindent{{\bf Definition 3.}} By the {\it minimal primary} of an
RCFT, we mean the unique chiral primary $o\in\Phi$ obeying (8) for all primaries $a\in\Phi$.
\medskip

\medskip\noindent{{\bf Fact 2.}} In any RCFT, {no primary can have smaller conformal
weight than that of the minimal primary} $o$.

\medskip To see this, let $0',0'',\ldots$ be the primaries
with minimal conformal weight. The argument of (7) shows that for
small $\epsilon>0$, the leading terms of $\chi_a(\epsilon\i)$ will be 
$$S_{a0'}\chi_{0'}(\i/\epsilon)+S_{a0''}\chi_{0''}(\i/\epsilon)+\cdots\ ,$$
that is to say
$$0\le S_{a0'}A_{0',0}+S_{a0''}A_{0'',0}+\cdots$$
for all $a\in\Phi$ (recall that $A_{b,n}$ are the coefficients of $\chi_b$).
By unitarity of $S$, and inequality (8), this forces $o$ to be one of the
primaries $0',0'',\ldots$ 

Every RCFT has a unique minimal primary. In a unitary RCFT, the minimal primary will be
the vacuum 0. In all cases known to this author, there is a unique primary with
minimal conformal weight (it does not seem to be known that this should
always be the case). Recall that the conformal weight is (essentially) the
energy, so it is tempting to identify the minimal primary with the true vacuum.
On the other hand, the true vacuum should be invariant under the Poincar\'e
group and in particular translations, yet typically the descendent $L_{-1}o$ will not vanish
(see the explicit example in \S6). We will find that in a nonunitary theory,
all of the familiar properties usually ascribed to the vacuum are now
distributed between 0 and $o$.

The {\it quantum-dimension} is the positive number $S_{ao}/S_{0o}$. It is the maximal
(Perron-Frobenius [34]) eigenvalue of the fusion matrix $N_a$, and so obeys
$$\eqalignno{S_{ao}\ge &\,S_{0o}\ , &(9a)\cr |S_{ao}\,S_{0b}|\ge
&\,|S_{0o}\,S_{ab}|\ ,&(9b)\cr N_{ab}^c\le {\rm min}&\{{S_{ao}\over
S_{0o}}, {S_{bo}\over S_{0o}}, {S_{co}\over S_{0o}}\}\ .&(9c)}$$
See e.g.\ [12] for proofs.  When there is a unique primary with minimal
conformal weight, then that primary of
course must be $o$, and the quantum-dimension $S_{ao}/S_{0o}$ equals
the limit ${\rm lim}_{q\rightarrow 1}\chi_a(q)/\chi_0(q)$.

A {\it simple-current} $j$ is by definition any primary with
quantum-dimension 1: i.e.\ $S_{j,o}=S_{0o}$. Any simple-current
corresponds to a permutation $J$ and a phase (`monodromy charge') $a\mapsto Q_J(a)\in \Q$ such that 
$N_{j,a}^b=\delta_{b,Ja}$, $J0=j$, and [35] 
$$S_{Ja,b}=e^{2\pi\i \,Q_J(b)}\,S_{ab}\ .\eqno(10a)$$
We'll usually identify the simple-current $j$ with the permutation $J$.
 From (10a) we get $Q_{JJ'}(a)\equiv Q_J(a)+Q_{J'}(a)$ (mod
1). Hence if $J^n=id.$, then $Q_J(a)\in {1\over n}\Z$. Also,
$Q_J(\si_\ell a)\equiv \ell \,Q_J(a)$ (mod 1) and
$\si_\ell(Ja)=J^\ell\si_\ell a$, for all Galois automorphisms
$\si_\ell$. Although (8) requires $Q_J(o)\in\Z$ for all
simple-currents $J$, one distinction with the more familiar unitary
theories is that {\it the monodromy charge $Q_J(0)$ of the vacuum can be a half-integer.} This introduces the
following modifications in the formulas applying to the unitary
theories: $N_{ab}^c\ne 0$ implies
$Q_J(a)+Q_J(b)\equiv Q_J(c)+Q_J(0)$ (mod 1), so
$Q_J(J'a)\equiv Q_J(J')+Q_J(a)+Q_J(0)$ (mod 1). Expanding $S_{J,J'}$
in two ways gives $Q_J(J')+Q_{J'}(0)\equiv Q_{J'}(J)+Q_J(0)$ (mod 1).
Finally (see [12] for the arguments)
$$\eqalignno{h_{Ja}-h_a\equiv&\, h_J+Q_J(0)-Q_J(a)\ ({\rm mod}\ 1)\ ,&(10b)\cr
2h_J\equiv&\,Q_J(0)-Q_J(J)\ ({\rm mod}\ 1)\ .&(10c)}$$
Note that for nonunitary theories, a simple-current is defined by
$S_{jo}=S_{0o}$, rather than the simpler but potentially weaker $S_{j0}=\pm
S_{00}$ (but see Consequence 2(vi) next section).

Both the vacuum 0 and the minimal primary $o$ obey $C0=0$
and $Co=o$, because their $S$-columns  are real.
Also, neither 0 nor $o$ can be fixed by any simple-current $J\ne id.$
(because otherwise $S_{a0}=0$ or $S_{ao}=0$ for any $a$ with
$Q_J(a)\not\in\Z$). 

There are arguments valid for unitary RCFTs which break down for
nonunitary ones, because the vacuum 0 and the minimal primary $o$ are
distinct. For instance, consider the proof that for a given choice of
{\it unitary} modular data, there are only a finite number of possible
modular invariants $M$. In particular, we get the bounds [36]
$$M_{ab}=|\sum_{c,d\in\Phi}S_{ac}M_{cd}S_{bd}^*|\le {S_{a0}\,S_{b0}\over
S_{00}^2}\sum_{c,d\in\Phi} S_{0c}M_{cd}S_{0d}= {S_{a0}\,S_{b0}\over
S_{00}^2}\ ,$$
using the triangle inequality, (9b), and the uniquess of the vacuum
$M_{00}=1$. For nonunitary modular data, this argument breaks down, as
it yields the bound $M_{ab}\le M_{oo}S_{ao}S_{bo}/S_{0o}^2$. All other
finiteness proofs for modular invariants known to this author,
similarly break
down for nonunitary modular data, because we don't have a bound on the
multiplicity $M_{oo}$.
Note however that this argument does give us the interesting fact that
in any RCFT $$M_{oo}\ge 1\ .$$

For example consider the nonunitary modular data
$$\eqalignno{S=&\,{1\over 6}\left(\matrix{1&1&2&2&2&2&-3&-3\cr 1&1&2&2&2&2&3&3
\cr 2&2&4&-2&-2&-2&0&0\cr 2&2&-2&4&-2&-2&0&0\cr 2&2&-2&-2&-2&4&0&0\cr
2&2&-2&-2&4&-2&0&0\cr -3&3&0&0&0&0&3&-3\cr -3&3&0&0&0&0&-3&3}\right)\ ,&(11a)\cr
T=&\,{\rm diag}(1,1,1,1,\xi_3,\xi_3^*,-1,1)\ .&}$$
Number the primaries from 0 to 7, and let the 0th one be the vacuum. Then 
the minimal primary $o$ is `1'. This data 
 has the same fusions as the quantum-double of the finite group $S_3$ (which is
 its `uniformisation'); although
$S$ obeys Fact 1, we know of no RCFT which realises it.
It has infinitely many modular invariants: for example, choose any integer
$m\ge 0$, then
$${\cal Z}= (\sum_{a=0}^7 |\chi_a|^2)+m\,|\chi_1+\chi_3+\chi_7|^2\eqno(11b)$$
defines a modular invariant. Here, $M_{oo}=m+1$ can be arbitrarily large.

The same thing was noticed before, in a related context: there are infinitely
many modular invariants for the admissible representations of
affine algebra $\widehat{{\rm su}}(2)$ at
fractional (nonintegral) level $k$ [37]. Why this didn't happen for the
nonunitary Virasoro minimal models, is explained by Fact 3 below. It
is somewhat surprising that, although there are infinite numbers of
$\widehat{{\rm su}}(2)$ modular invariants at each fractional level,
there are only a finite number of modular invariants for their cosets
(namely the nonunitary minimal models) --- after all we tend to think
that a source of coset modular invariants are products of the modular
invariants of the component affine algebras. Of course the explanation,
ultimately, is field identification.

  Another useful fact about modular invariants in unitary modular
  data, is that for any simple-currents $j,j'\in\Phi$, $M_{j,j'}=0$ or
  1, and $M_{j,j'}=1$ iff the following selection rule holds for all $a,b\in\Phi$:
$$M_{ab}\ne 0\qquad\Longrightarrow \qquad Q_J(a)\equiv Q_{J'}(b)\
({\rm mod}\ 1)\ .\eqno(12)$$
The proof of this also collapses for nonunitary RCFTs, for the same reason; the argument instead tells us
that
$M_{Jo,J'o}\le M_{oo}$ with equality only if the above selection rule holds.

Similar remarks are valid for the cylindrical partition functions (i.e.\ the
{\smcap nim}-reps).
Any {\smcap nim}-rep is uniquely decomposable into a direct sum of irreducible {\smcap nim}-reps.
For any {\smcap nim}-rep $n_a$ of unitary modular data, the multiplicity
mult$_n(0)$ of the vacuum in (4)
precisely equals the number of irreducible {\smcap nim}-reps which compose
$n_a$. Each of these irreducible summands would correspond to a family of boundary
conditions which completely decouples from the other families.
The uniqueness of the vacuum then tells us that the {\smcap nim}-rep of
any unitary RCFT must be indecomposable [24]. The finiteness result
for {\smcap nim}-reps of unitary modular data follows from this (see [24] for details).
For {\it
nonunitary} modular data, the number of irreducible {\smcap nim}-reps composing a
given {\smcap nim}-rep equals the multiplicity mult$_n(o)$ of the minimal primary $o$.  
A priori, a {\smcap nim}-rep corresponding to a nonunitary RCFT may be
decomposable; hence the finiteness result [24] for {\smcap nim}-reps breaks down unless
we know that the multiplicity of $o$ is bounded.

\bigskip\noindent{{\bf 4. The Galois Shuffle and consequences}}\medskip

For the reasons given last section, it is important to answer the following:

\medskip\noindent{{\bf Question.}} What is the multiplicity
mult$_n(o)=M_{oo}$ of the minimal primary?\medskip

The Virasoro minimal models were classified in [7]; checking the
answer there, we get multiplicity $M_{oo}=1$ for all of them. We can explain
and generalise this result:

\medskip\noindent{{\bf Fact 3.}} In  any $W_N$ $(p,p')$ minimal 
model, for any positive integers $N,p,p'$, modular invariance of the 1-loop partition function forces
the multiplicity mult$_n(o)=M_{oo}$ of the minimal primary to be 1.

\medskip We prove this in \S6. Fact 3 is new.

\medskip\noindent{{\bf Conjecture.}}  The multiplicity
mult$_n(o)=M_{oo}$ of the minimal primary in any RCFT is 1.\medskip

This makes some sense, as there are reasons for thinking of the minimal primary as the true
vacuum\footnote{$^3$}{{\smal Nevertheless, as pointed out to the author by
Philippe Ruelle, curious things can happen in even the nicest unitary RCFTs.
For example, the Ising model is unitary, but it has nonunitary observables,
according to [38]. That paper also suggests that the conformal weight 0 primary there
need not be the identity field. In [39] we see that the 3-state Potts model on the
cylinder can have a degenerate vacuum, for some choices of boundary conditions.}}.
It can also (see Consequence 2(ii)) follow from the Galois Shuffle property
given shortly. Collecting results from last section, we get

\medskip\noindent{{\bf Consequence 1.}} Suppose the multiplicity
mult$_n(o)=M_{oo}$ of the minimal primary in a given RCFT is 1. Then:

\item{{\bf 1(i)}} There will be only finitely many modular invariants $M$ with the given
(chiral) modular data --- in fact we get the inequality $M_{ab}\le
S_{ao}S_{bo}/S_{0o}^2$;

\item{{\bf 1(ii)}} Any {\smcap nim}-rep corresponding to a modular invariant $M$ will be
indecomposable. There are only finitely many indecomposable {\smcap nim}-reps.

\medskip In a superficially independent direction, further consequences occur when we assume more.

\medskip\noindent{{\bf Definition 4.}} We say an RCFT has the {\it GS} (or {\it
Galois shuffle}) property, if there is a simple-current
$J_o$ (possibly the identity), and a Galois automorphism $\si_o$
(possibly the identity),
such that the minimal primary $o$ in the theory is related to the vacuum by
$o=J_o\si_o 0$.\medskip

Of course this is trivially obeyed by any {\it unitary} RCFT. We verify in 
\S6 that:

\medskip\noindent{{\bf Fact 4.}} All $W_N$ minimal models possess the
GS property.\medskip

The same is true for the minimal WSO($2N)$ models, but we won't prove it
here. The modular data for the minimal W$\overline{\frak g}$ models, for
$\overline{\frak g}$ nonsimply laced, is a little more complicated and we haven't
looked at them yet.

Modular data may not obey this property. For example, the 1-dimensional modular
data $S=(-1), T=(-1)$ does not obey it. On the other hand, that 1-dimensional
modular data violates Fact 1 and so can't be realised by an RCFT. A more
serious example is the tensor product of $\widehat{G_2}{}_{,1}$ 
 with the Yang-Lee model. That this doesn't have the
GS property can be seen from Consequence 2(iv) below: this model has central charge
$c=-8/5$, and four primaries, with conformal weights $0,\pm 1/5,2/5$, so
$h_o=-1/5$, $h_J=0$ and  the congruence in 2(iv) reduces to $\ell^2\equiv -2$
(mod 15), which has no solutions.
More generally, the tensor product of a nonunitary model with its unitarisation (defined
below)  typically won't satisfy the GS property.  

Even though the GS property does not hold for all RCFTs, {\it when} it holds
for a given RCFT (which is often, as we'll see) it has several consequences.
We begin by listing some of its immediate consequences, without assuming
the Conjecture. The most important of these are Consequences 2(ii), (iv), (v) and (viii)
--- in particular, (iv) reduces to a statement about quadratic residues and 
becomes
a stronger constraint the more distict primes divide the order of $T$.
Consequences
2(i),(iii),(iv) can be interpreted as constraining the possible simple-currents
$J_o$ and Galois automorphism $\si_o$. Consequences 2(vi),(vii) are of technical interest.

\medskip\noindent{{\bf Consequence 2.}} Let 0 be the (chiral) vacuum and $o$ the minimal
primary. Suppose $o=J_o\si_o 0$ for some Galois automorphism $\si_o$ and some
simple-current $J_o$. Then:

\smallskip\item{{\bf 2(i)}} The simple-current $J_o$ has order 1 or 2;

\item{{\bf 2(ii)}} If $J_0=id$, then the multiplicity mult$_n(o)=M_{oo}$
of the minimal primary must be 1 (i.e.\ the Conjecture must hold for this RCFT).

\item{{\bf 2(iii)}} For any simple-current $J$, $Q_J(J_o)\in\Z$
and $Q_{J_o}(J)\equiv Q_{J_o}(0)+Q_J(0)$ (mod 1).

\item{{\bf 2(iv)}} The conformal weight $h_o$ of the minimal primary
satisfies $$h_o\equiv {c\over
24}\,(1-\ell^2)+h_J\ ({\rm mod}\ 1)$$
for some integer $\ell$ coprime to the denominator of $c/24$, and for
some simple-current $J$ of order 1 or 2 (so $4h_J\in \Z$).\medskip  

\item{{\bf 2(v)}} The fusion rules are exactly reproduced by
modular data $\widehat{S},\widehat{T}$, which is {\it unitary} in the sense of Def.2 (this
{\it unitarisation} is explicitly given in equation (13) below).

\item{{\bf 2(vi)}} A primary $j\in \Phi$ will be a simple-current,
iff $S_{j0}=\pm S_{00}$.

\item{{\bf 2(vii)}} $|S_{a0}|\ge S_{o0}$.

\item{{\bf 2(viii)}} The list of modular invariants $M$ of $S,T$ with
$M_{oo}=1$ is identical to that of the modular invariants $\widehat{M}$ of
the unitarisation $\widehat{S},\widehat{T}$ which obey
$\widehat{M}_{J_o,J_o}=1$; the (indecomposable) {\smcap nim}-reps of $S,T$ can be
identified with those of the unitarisation
$\widehat{S},\widehat{T}$.\medskip

First note that (1c), $C\si=\si C$, and $C0=0$, tell us that each
column $S_{\updownarrow,\si 0}$ is real, for any Galois $\si$. Hence
Consequence 2(i) follows from (10a), because both columns
$S_{\updownarrow, \si_o0}=S_{\updownarrow,J_o^{-1}o}$ and $S_{\updownarrow o}$ are
real. Consequence 2(ii) follows from (5c).
To see
Consequence 2(iii), note that the inequality $S_{J,o}>0$ requires
$Q_J(J_o\si_o0)\in\Z$. 
In Consequence 2(iv), take $J=J_o$ and $\si_\ell=\si_o$, and use (5b)
and (10b).
Since $S_{j0}=\pm S_{00}$ iff $S_{j,J_o}=\pm S_{0,J_o}$
iff $\si_o(S_{j,J_o})=\pm \si_o(S_{0,J_o})$, we get 2(vi). Applying
$\si_o\circ \si_o^{-1}=id.$ to $S_{a0}$ and using (9a), gives 2(vii).
Since the fusions of $S$ and $\widehat{S}$ are identical (as was discussed
after eqs.(6)), so are
their {\smcap nim}-reps. The statement in 2(viii) about modular invariants
follows from (5c): an integral matrix $M$ commutes with $S$ and $T$ iff
the integral matrix $\widehat{M}:=PMP^t$ commutes with $\widehat{S}$ and $\widehat{T}$.

The unitarisation $\widehat{S},\widehat{T}$ mentioned in Consequence 2(v) is
given explicitly as follows. Recall equations (6a), (6b).
Put $J=J_o$ and $\sigma=\sigma_o$. Define the following matrices $$\eqalignno{
\widehat{S}_{ab}:=\epsilon_\si(0)\,\si(S_{Ja,Jb})=&\,\epsilon_\si(0)\,(-1)^{Q_J(a)+Q_J(b)+Q_J(0)}\,\si(S_{ab})\ ,&(13a)\cr
\widehat{T}_{ab}:=&\,\delta_{ab}\,\epsilon_\si(0)\,(T_{Ja,Jb})^\ell\ ,&(13b)}$$
where $\ell\in\Z$ corresponds to $\si$, i.e.\ $\sigma$ acts on roots of unity
by $\si(\xi)=\xi^\ell$.
Requiring the vacuum 0 in the unitarised modular data to have
conformal weight 0, we find that the unitarisation
$\widehat{S},\widehat{T}$ has central charge
$$\widehat{c}\equiv \ell c-24\ell h_{J_o}+12\epsilon\qquad ({\rm mod}\
24)\ ,$$
where $(-1)^\epsilon:=\epsilon_\sigma(0)$.

As explained after (6b), the unitarisation obeys all of the conditions stated
in Definitions 1 and 2, and an easy Galois argument shows that it automatically
obeys the additional conditions (3) and identities given in [14] (provided $S,T$ do). 
It is conceivable however that $\widehat{S},\widehat{T}$ may not always be
realisable by an RCFT --- for instance $\widehat{S}$ could conceivably
violate {Fact 1}. Nevertheless, in the few cases this author has been able
to check, the uniformisation always seems to be realised by a completely
healthy unitary RCFT. We return to this in the concluding section.

Although the minimal primary $o$ is uniquely determined from the
modular data, the simple-current $J_o$ and Galois automorphism $\si_o$ may not
be unique. {\it There can be more than one unitarisation of nonunitary modular data.}
In particular, let $G_0$ be the set of all Galois
automorphisms $\si$ such that $\si(S_{00})=\pm S_{00}$. Then $\si_o$
will be unique only up to $G_0$:  $\si_o$ can be replaced by
any other $\si$ in its coset $G_0\si_o$; but once the Galois automorphism
$\si_o$ has been chosen, the simple-current $J_o$ will be uniquely determined.

Further results hold if we assume both the Conjecture and the GS property.

\medskip\noindent{{\bf Consequence 3.}} Let 0 be the (chiral) vacuum and $o$ the minimal
primary. Suppose $o=J_o\si_o 0$ for some Galois automorphism $\si_o$ and some
simple-current $J_o$. Assume that the multiplicity mult$_n(o)= M_{oo}$ in the full RCFT
is 1. Then:

\item{{\bf 3(i)}} The list of possible toroidal and cylindical partition functions
are identical for $S,T$ as for its unitarisation $\widehat{S},\widehat{T}$.

\item{{\bf 3(ii)}} The multiplicity mult$_n(J_o)=M_{J_o,J_o}=1$.

\item{{\bf 3(iii)}} We have the
symmetry $M_{J_oa,J_ob}=M_{ab}$.

\item{{\bf 3(iv)}} We have the selection rule $M_{ab}\ne
0\ \Rightarrow\ Q_{J_o}(a)\equiv Q_{J_o}(b)$ (mod 1).

\item{{\bf 3(v)}} If $J_o\ne id$, the {\smcap nim}-rep $n$ is 2-colourable.

\medskip First let us define 2-colourability of the {\smcap nim}-rep.
Let ${\cal B}$ be the set labelling the rows and columns of the
{\smcap nim}-rep (i.e.\ the vertices of the fusion graphs, equivalently the
labels of the boundary states).  We can assign each $x\in {\cal B}$ to a number
$q_x=0,{1\over 2}$ such that the selection rule
$$n_{ax}^y\ne 0\ \Rightarrow\ Q_{J_o}(a)+q_x-q_y\in\Z$$
holds for all $a\in\Phi$, $x,y\in{\cal B}$. For example, the {\smcap nim}-reps
of the Virasoro minimal models, at least those
compatible with the modular invariants, can be found in [40]. It can be
verified explicitly that all of them are 2-colourable. The {\smcap nim}-reps 
built out
of tadpoles aren't 2-colourable, but they also aren't compatible with any
modular invariants.

Consequence 3 is now easy to prove. 3(i) is just 2(viii). The Galois symmetry
(5c) with $\sigma=\sigma_o$ tells us that
$1=M_{00}=M_{J_oo,J_oo}$. Because $M_{J_oo,J_oo}=M_{oo}$, we get 3(ii),
and the rest automatically follows.

Consequence 3(i) reduces the classification of nonunitary models to that
of the more familiar unitary ones. 
In \S6 we give the unitarisation of the $W_N$ minimal models, and we find
for instance that the affine algebra $\widehat{{\rm su}}(3)$ gives the
uniformisation of certain $W_3$ models. We thus obtain for free the
classification of those $W_3$ models, in the bulk. With a little additional work, the
classification of all nonunitary $W_3$ minimal models (in the bulk) can be 
obtained [41].

\medskip\noindent{{\bf Consequence 4.}} The complete list of $W_3$ minimal
models at $(p,4)$ is in exact one-to-one correspondence with the list of $\widehat{{\rm su}}
(3)$ level $p-3$ modular invariants. In particular, there are precisely four
modular invariants for each value of $p$, except for $p=3$ (where there is only
one) and $p=5$ (where there are only two). Those four modular invariants  are
all constructed using simple-currents
and/or charge-conjugation in the standard way. Each of those modular invariants
has a compatible {\smcap nim}-rep.\medskip

Of course $p$ there must be odd (since it must be coprime to 4). It is important
to note that Consequence 4 is a theorem independent of the general validity
of the conjecture, because Facts 3 and 4 tell us that the minimal $W_3$
models obey all the hypotheses of Consequence 3. The proof of Consequence 4,
and the explicit correspondence between the $W_3$ modular invariants and
those of $\widehat{{\rm su}}(3)$, will be given at the end of \S6.

\bigskip\noindent{{\bf 5. Comments on affine algebras at fractional level}}\medskip

Let $\g=X_r^{(1)}$ be any affine algebra.
For the relevant facts and notation about integrable representations of $\g$,
and corresponding SL$_2(\Z)$ representations, see e.g.\ [4].

The so-called {\it admissible representations} of $\g$ at
fractional level $k=t/u$ [6] share many properties with the better
known integrable representations. Their importance for RCFTs, and this
paper, lies in the quantum Drinfeld-Sokolov reduction method for constructing 
W-algebras (see [8] for a review)
and their parallel use in GKO cosets for nonunitary models. But a natural
question, which has received much attention (see e.g.\ [42] and
references therein), is: Is there an RCFT
which corresponds more directly to the admissible representations,
roughly in the way that the integrable data is directly realised by
the Wess-Zumino-Witten models [3]? As is well-known, naively placing the
admissible $S$ matrix $S^{{\rm adm}}$ into Verlinde's formula fails to
produce nonnegative integer fusions (indeed a general though
conjectural expression for these `fusions', manifestly demonstrating that they
can be integers of either sign, is given in [43]). This means that the desired correspondence is
more subtle, if it exists at all. Indeed, for $\widehat{{\rm su}}$(2) the generally
accepted wisdom of [44,45] has now been challenged by [46] and independently by
[47]. For example, [47] suggests that  a CFT corresponding to admissible
fusions will be both nonunitary and quasi-rational.

In this section we make a couple of preliminary remarks, relating Galois
to the admissible $S$ matrix $S^{{\rm adm}}$. This underlies our
calculations in \S6, and (we believe) lends support to our expectation that the
GS property will be quite common among RCFT.

Let $g$ be the dual Coxeter number of $\overline{\g}=X_r$.
Write the level $k=t/u$, where gcd$(t,u)=1$, $t\in\Z$,
$u\in\{1,2,3,\ldots\}$. Put $k^I=u\,(k+g)-g$ and $k^F=u-1$, for
reasons to be clear soon. The integrable case is recovered when the
denominator $u=1$. Admissibility requires $k\ge ({1\over u}-1)g$, and we will
assume for most of  this
section an additional coprimeness condition: for $\overline{\g}=\,$su($N$), the
denominator
$u$ must be coprime to $N$; for $\overline{\g}=\,$so$(N)$, sp($N$), $F_4$, or  $E_7$, $u$
must be odd; and for $\overline{\g}=E_6$ or $G_2$, $u$ must not be a multiple of 3. This
coprimeness condition is necessary for the direct Galois
interpretation to be given shortly; when it fails the Galois
interpretation is slightly more subtle (see the discussion on [48]
when $u$ is even, given at the end of this section).

The admissible highest weights $P^k_{{\rm adm}}$ consist of pairs
$\la=(\la^I,\la^F)$, where $\la^I\in P_+^{k^I}$ is an integrable
highest weight of $\g$ of level $k^I\in\Z_+$, and where $\la^F$ belongs to a
certain finite set of level $k^F\in\Z_+$ weights of $\g$. For instance, for
$\widehat{{\rm su}}$(2), $\la^F\in P_+^{k^F}$, so $P^k_{{\rm adm}}=P_+^{k^I}\times
P^{k^F}_+$. For $\widehat{{\rm su}}$(3), $\la^F$ belongs to this {\it disjoint} union of
$P_+^{k^F}$ with the subset of $P_+^{k^F}$ with first Dynkin label
$\la_1^F\ge 1$. In general, the cardinality of $P_{{\rm adm}}^+$ will be
$\|P_+^{k^I}\|\,u^r$ where $r$ is the rank of the Lie algebra $\overline{\frak g}$.
The modular $S$ matrix $S^{{\rm adm}}$ is given by [6]
$$\eqalignno{S^{{\rm adm}}_{\la\mu}=&\,\pm
F_m\exp[2\pi\i\,\{(\la^I+\rho|\mu^F)+(\la^F|\mu^I+\rho)-(k+g)(\la^F|\mu^F)\}]\,&\cr&\times\sum_{w\in
W}{\rm det}(w)\exp[{-2\pi\i\over
k+g}(w(\la^I+\rho)|\mu^I+\rho)]=\varphi_{\la\mu}S^{(k)}_{\la^I\mu^I}\ ,&(14)}$$
where $F_m$ is some constant (independent of $\la,\mu$), and the sign
$\pm 1$ depends on $\la,\mu$ (it is given explicitly in e.g.\ eq.(2.46) of [49]).
$W$ here is the (finite) Weyl group of $\widehat{\g}$. The modulus $|\varphi_{\la\mu}|=u^{-r/2}$ is
constant, and the matrix $S^{(k)}$ is the sum over the Weyl group $W$,
normalised by $u^{r/2}F_m$.

$S^{{\rm adm}}$ is unitary and symmetric, but $S$ does not have a column of
constant phase, and therefore Verlinde's formula (2) will yield
negative `fusions' $N_{\la\mu}^\nu$. So the admissible modular
matrices $S^{{\rm adm}},T^{{\rm adm}}$ don't constitute modular data (although
they define a representation of SL$_2(\Z)$ as before). 
 In modular data, the charge-conjugation
matrix $C=S^2$ is a permutation matrix; for the admissible data,
$C=S^2$ is a {\it signed} permutation matrix. Typically,
there won't be a {\it unique} admissible weight $\la\in P^k_{{\rm adm}}$
with minimal conformal weight.

Returning to formula (14), our observation in this section is simply that, up to a
sign independent of $\la,\mu$, $S^{(k)}_{\la^I\mu^I}=\pm
\si(S^{(k^I)}_{\la^I\mu^I})$, where $\si$ is a Galois automorphism
depending on $u$ but not $\la,\mu$, and where $S^{(k^I)}$ is the
integrable `WZW' $S$ matrix for ${\frak g}$ at integer level $k^I$.

What this means is that the $\la^I$ part of admissible representations
is understandable. If for instance we adjusted the phases $\varphi\rightarrow
\varphi'$ in (14)
appropriately, the admissible $S$ matrix $S^{{\rm adm}}$ would become a
factorised matrix $S^{{\rm fac}}=\varphi'\otimes S^{(k)}$ which, when placed in Verlinde's formula
(2), would yield the factorised nonnegative integer fusions $N'_{\la^F\mu^F}{}^{\nu^F}\,N^{(k^I)\
\nu^I}_{\la^I\mu^I}$ where $N^{(k^I)}$ are the
usual integrable WZW fusions for ${\frak g}$ at level $k^I$, and where the
fractional part $N'$ has simple-current fusions.

For example,  for $\widehat{{\rm su}}$(2) at level $k=t/u$, write
$\la=(\la^I,\la^F)=:(l,l')$, $\mu=:(m,m')$, where $\la^I\in
P_+^{t+2u-2}({\rm su}(2))=\{0,1,\ldots,t+2u-2\}$ and
$\la^F\in\{0,1,\ldots,2u-1\}$. Then
$$S^{{\rm adm}}_{\la\mu}=\sqrt{{2\over
u^2\,(k+2)}}(-1)^{l'\,(m+1)+m'\,(l+1)}e^{-\pi\i t
l'm'/u}\sin({\pi\,(l+1)(m+1)\over k+2})\ .$$
If we were to drop the $(-1)^{l'\,(m+1)+m'\,(l+1)}$ factor, then the
resulting factorised $S$ matrix $S^{{\rm fac}}$ would yield the factorised fusions
$N'_{l',m'}{}^{n'}\,N^{(t+2u-2)\ n}_{lm}$ where 
$N'_{l',m'}{}^{n'}=\delta_{n',l'+m'\ ({\rm mod}\ 2u) }$ and $N^{(t+2u-2)}$
are $\widehat{{\rm su}}(2)_{t+2u-2}$ fusions. 
 The Galois automorphism $\si$ involved here
corresponds to any $\ell$ coprime to $u$ satisfying $\ell\equiv u$
(mod $8(t+2u)$).

A more subtle idea along these lines was suggested  in [48], who in
effect changed the sign $(-1)^{l'\,(m+1)+m'\,(l+1)}$ to $(-1)^{\ell'\,m+m'\,\ell}$
(for $u$ odd) and to $(-1)^{(\ell'-1)\,(m+1)+(m'-1)\,(\ell+1)}$ (for
$u$ even). The resulting fusions are again in the factorised
form: $\widehat{{\rm su}}(2)_{t+2u-2}$ fusions times $\Z_{2u}$ fusions. Although 
these fusions differ from those in e.g.\ [44,45], they are
consistent with the in-depth analysis of [46] for $\widehat{{\rm su}}(2)$ at level $k=-4/3$.
Incidentally, this modular data of [48] is nonunitary; its
unitarisation corresponds to the Galois automorphism with $\ell$ odd and
congruent to $u$ (mod $t+2u$). 

What value if any the observation in this section has, is unclear at this point. However
it is implicit in the following section.

\bigskip\noindent{{\bf 6. Examples}}\medskip

The classic examples of  nonunitary RCFTs come from the Virasoro minimal models.
Let $p>p'\ge 2$ be coprime integers. 
The primary fields $\varphi_{rs}$ here are parametrised by all pairs
$(r,s)$ where $1\le r\le p'-1$, $1\le s\le p-1$, and $p's<pr$. So the number
 of primaries is $(p-1)(p'-1)/2$. The conformal
weight $h_{rs}$ of $\varphi_{rs}$ is
$$h_{rs}={(pr-p's)^2-(p-p')^2\over 4pp'}\ .\eqno(15a)$$
The $S$ matrix has entries
$$S_{rs;r's'}=2\sqrt{2\over pp'}(-1)^{1+sr'+rs'}\sin(\pi {p\over p'}rr')
\sin(\pi {p'\over p}ss')\ .\eqno(15b)$$
The central charge of the $(p,p')$ minimal model is $c=1-6\,(p-p')^2/pp'$.
It is unitary iff $p-p'=1$.
The vacuum 0 corresponds to primary $(r,s)=(1,1)$. The simple-current
is $j=(p'-1,1)$, corresponding to permutation $J(r,s)=(p'-r,s)$ (if
$pr+p's<pp'$) or $(r,p-s)$ (otherwise); the monodromy charge in (10a) is
$Q(r,s)=(1+pr+p's)/2$. There is a unique primary
$o=(r_o,s_o)$ with minimal conformal weight, corresponding to the unique $(r,s)\in\Phi$
obeying $pr-p's=1$. 

The Galois group here can be taken to be $\Z_{8pp'}^\times$. When $r_o$ and $s_o$ are
both odd, choose any $\ell$ obeying $\ell\equiv r_o$ (mod $2p'$) and
$\ell\equiv s_o$ (mod $2p$), and take $J_o=(1,1)$. When instead $r_o$ is even, so
both $s_o$ and $p'$ will be odd, choose $\ell\equiv p'-r_o$ (mod $2p'$) and
$\ell\equiv s_o$ (mod $2p$), and $J_o=(p'-1,1)$. Finally, 
if $s_o$ is even, then
both $r_o$ and $p$ will be odd, and choose $\ell\equiv r_o$ (mod $2p'$) and
$\ell\equiv p-s_o$ (mod $2p$), and again $J_o=(p'-1,1)$. In all cases,
by the Chinese remainder theorem, there's a unique solution in the
range $1\le \ell\le 2pp'$; it will lie in $\Z_{8pp'}^\times$ (i.e.\ be coprime
to 2, $p$ and $p'$), so let
$\si_o$ be the corresponding Galois automorphism. Then
$o=J_o\si_o(1,1)$, showing that the GS property holds here.

For a concrete example, consider the (7,3) minimal model $(c=-25/7$), with primaries
$$\{(1,1),(1,2),(2,1),(2,2),(2,3),(2,4)\}$$ 
taken in that order. Then
$$\eqalign{S_{rs;r's'}=&\,\sqrt{{2\over 7}}\,\left(\matrix{-d&a&d&-a&-b&b\cr
a&b&a&b&d&d\cr d&a&d&a&-b&-b\cr -a&b&a&-b&d&-d\cr -b&d&-b&d&-a&-a\cr
b&d&-b&-d&-a&a}\right)\ ,\cr T_{rs;r's'}={\rm
diag}\{\exp[\pi\i{25\over 84}],&\exp[\pi\i{-5\over 84}],\exp[\pi\i{235\over 84}],
\exp[\pi\i{121\over 84}],\exp[\pi\i{43\over 84}],\exp[\pi\i{1\over 84}]\}\ ,}$$
where $a=\sin(\pi/7)\approx 0.434$, $b=\sin(2\pi/7)\approx 0.782$, and $d=\sin(3\pi/7)\approx 0.975$.
The positive column (namely the second) corresponds to the minimal
primary $o=(1,2)$ (note that $7\cdot 1
-3\cdot 2=1$). The Galois automorphism $\si_o$ corresponds to
$\ell=19$, and the simple-current $J_o=(2,1)$ is also nontrivial.
The unitarisation is
$$\eqalign{\widehat{S}=&\,\sqrt{2\over 7}\,\left(\matrix{a&b&a&b&d&d\cr b&d&-b&-d&-a&a
\cr a&-b&-a&b&-d&d\cr b&-d&b&-d&a&a\cr d&-a&-d&a&b&-b\cr d&a&d&a&-b&-b}\right)\ ,\cr
\widehat{T}={\rm diag}\{\exp[\pi\i{97\over 84}],&\exp[\pi\i{-53\over 84}],
\exp[\pi\i{-29\over 84}],\exp[\pi\i{73\over 84}],\exp[\pi\i{19\over 84}],\exp[\pi\i{-23\over 84}]\}\ .}$$
Incidentally, this unitarisation is the modular data for affine algebra
$\widehat{{\rm su}(2)\oplus E_8}$ at level (5,1); more generally, the $(p,3)$
Virasoro minimal model has unitarisation corresponding to $\widehat{{\rm su}}(2)$
at level $p-2$, whenever $p$ is odd (ignoring a certain number of $\widehat{E_{8}}{}_{,1}$'s,
whose only purpose is to adjust $c$ by an appropriate multiple of 8).

The characters $\chi_{rs}$ of the minimal models have been explicitly calculated
[50]: for example, for the (7,3) model the characters of the vacuum and minimal
primaries are
$$\eqalign{\chi_0=\chi_{11}=&\,q^{25/168}\,(1+q^2+q^3+2q^4+\cdots)\ ,\cr
\chi_o=\chi_{12}=&\,q^{-5/168}\,(1+q+q^2+2q^3+\cdots)\ .}$$
In particular we see that, although the translation operator $L_{-1}$ kills
the vacuum 0, it doesn't kill the minimal primary $o$. This illustrates one
of the ways the minimal primary doesn't behave like a vacuum.

More generally, we can consider the minimal $W_N$  models 
(cf.\ [51,52,49,43]). These theories are also parametrised
by a pair $p>p'\ge N$ of coprime integers. As before, they are unitary iff $p=p'+1$.
The primaries consist of all $(J,J)$-orbits $(J^i\la,J^i\mu)$, where $\la\in P_{++}^{p'},
\mu\in P_{++}^p$ --- by `$P_{++}^m$' we mean all $N$-tuples $\nu_i$ of positive
integers, with $\sum_{i=0}^{N-1}\nu_i=m$. Algebraically, $P_{++}^m$ are the
level $m-N$ highest weights of $\widehat{{\rm su}}(N)$, shifted by the Weyl
vector $\rho=(1,1,\ldots,1)$. The simple-current $J$ of $\widehat{{\rm su}}(N)$
takes weight $\nu=(\nu_0,\nu_1,\ldots,\nu_{N-1})\in P_{++}^m$ to $(\nu_{N-1},\nu_0,\nu_1,\ldots,
\nu_{N-2})\in P_{++}^m$. So there are a total of $\bigl({p\atop N-1}\bigr)
\bigl({p'\atop N-1}\bigr)/N$ primaries. The matrix $T$ is given by the
conformal weights
$$h_{\la\mu}\equiv {|p\la-p'\mu|^2-(p-p')^2(N-1)N(N+1)/12\over 2pp'}\qquad
({\rm mod}\ 1)\eqno(16a)$$
and central charge $c=(N-1)\{1-N(N+1){(p-p')^2\over pp'}\}$.
The $S$ matrix is
$$S_{(\la,\mu)(\la',\mu')}=\alpha\,\exp(-2\pi\i\,[t(\la)\,t(\mu')+t(\mu)\,
t(\la')]/N)\,S^{(p/p')}_{\la,\la'}\,
S^{(p'/p)}_{\mu,\mu'}\ ,\eqno(16b)$$
where $\alpha$ is some irrelevant constant, where
$t(\la)=\sum_ii\la_i$, and where $S^{(m)}$ here 
is the usual (`integrable') affine $S$ matrix, expressed as usual as an alternating
sum over the (finite) Weyl group (see e.g.\ eq.(14.57) of [4]), formally
evaluated at (fractional)
level $m-N$. The vacuum 0 corresponds to the $(J,J)$-orbit containing
$(\rho,\rho)$, or more precisely $((p'-N+1,1,\ldots,1),(p-N+1,1,\ldots,1))$
(recall that $(k+1,1,\ldots,1)\in P_{++}^{k+N}$ corresponds to the vacuum in 
the $\widehat{{\rm su}}(N)_k$ WZW theory, and projects to the Weyl vector $\rho$ of su$(N)$).
An order-$N$ simple-current of this minimal model is
$(J,id)$ (or rather its ($J,J)$-orbit); it has monodromy charge $Q(\la,\mu)
\equiv{p\,t(\la)-p'\,t(\mu)\over N}+{N-1\over 2}$ (mod 1). As we will see below,
it generates all of the  simple-currents of the model.

The Galois group here can be taken to be the multiplicative group
$\Z_{4Npp'}^\times$. Choose any odd integer $\ell$, coprime to $N$,
obeying the congruences $\ell p\equiv 1$ (mod $p'$) and $\ell p'\equiv 1$ (mod $p$)
(this is always possible, by the Chinese remainder theorem and the fact that
gcd$(p,p')=1$). Let $\sigma_\ell$ be the
Galois automorphism corresponding to $\ell$, and write $(o',o'')\in P_{++}^p
\times P_{++}^{p'}$ for the image $\si_\ell 0$ of the vacuum under $\sigma_\ell$. We claim
that $(J^io',o'')$ is the minimal primary $o$, for some $i$.

To see this, first note that for any primaries $(\la,\mu)$, $(\la',\mu')$, we have
$$\bigl|{S_{(\la,\mu)(\la',\mu')}\over S_{0(\la',\mu')}}\bigr|=
\bigl|{S^{(p/p')}_{\la\la'}\over S^{(p/p')}_{\rho\la'}}\bigr|\,\bigl|
{S^{(p'/p)}_{\mu\mu'}\over S^{(p'/p)}_{\rho\mu'}}\bigr|=\bigl|\sigma^{-1}_\ell
{S^{(p)}_{\la\la'}\over S^{(p)}_{\rho\la'}}\bigr|\,\bigl|\sigma^{-1}_\ell
{S^{(p')}_{\mu\mu'}\over S^{(p')}_{\rho\mu'}}\bigr|
=\bigl| {S^{(p)}_{\la\la''}\over S^{(p)}_{\rho\la''}}\bigr|\,\bigl|
{S^{(p')}_{\mu\mu''}\over S^{(p')}_{\rho\mu''}}\bigr|\ ,$$
where $\sigma^{-1}_\ell(\la',\mu')=(\la'',\mu'')$.
Implicit in this calculation is that we can write $S^{(p)}_{\lambda\lambda'}=\psi_{
\lambda\lambda'}S'_{\lambda\lambda'}$ (similarly for $S^{(p')}_{\mu\mu'}$),
where $\psi_{\lambda\lambda'}=\exp[2\pi\i\,t(\lambda+\rho)\,t(\lambda'+\rho)/
pN]$ is a root of unity, and $S'_{\lambda\lambda'}$ is in the number
field $\Q[\exp[2\pi\i/p]]$. This means that $|\sigma_mS^{(p)}_{\lambda\lambda'}|
=|\sigma_mS'_{\lambda\lambda'}|$ depends only on the value of $m$ (mod $p$).
The maximum value of the right-side is the product of WZW quantum dimensions,
achieved by the vacuum $(\la'',\mu'')=0$ (recall (9b)).
In particular, the choice $(\la',\mu')=(o',o'')$ achieves this maximum, for all
primaries $(\la,\mu)$, as must the minimal primary $(\la',\mu')=o$ (again by
(9b)). Thus, for every primary $(\la,\mu)$ there is a phase $\varphi_{\la\mu}$
such that
$S_{(\la,\mu)(o',o'')}/S_{0(o',o'')}=\varphi_{\la\mu}
S_{(\la,\mu),o}/S_{0,o}$. Taking the absolute value of
$${S_{(\la,\mu)(o',o'')}\over S_{0(o',o'')}}
{S_{(\la',\mu')(o',o'')}\over S_{0(o',o'')}}=
\sum_{(\la'',\mu'')} N_{(\la,\mu)(\la',\mu')}^{(\la'',\mu'')}
{S_{(\la'',\mu'')(o',o'')}\over S_{0(o',o'')}}$$
and using the triangle inequality, we find that the assignment
$(\la,\mu)\mapsto \varphi_{\la\mu}$ defines a grading on the $W_N$
fusion ring at $(p,p')$. Hence $(\la,\mu)\mapsto \varphi_{\la\mu}S_{(\la,\mu)0}/S_{00}$
defines a 1-dimensional representation of the fusion ring, and so
$$\varphi_{\la\mu}{S_{(\la,\mu)0}\over S_{00}}
={S_{(\la,\mu),A}\over S_{0,A}}\eqno(17)$$
for some primary $A$. We want to show $A$ is a simple-current, i.e.\ that
$S_{A,o}=S_{0,o}$. By unitarity of the matrix $S$, the norm
 of the $0$-column must equal that of the $A$-th column,
and so (17) implies $|S_{00}|=|S_{0,A}|$.
Substituting $o$ for $(\la,\mu)$ in (17) and using positivity (8)
now concludes the proof of Fact 4: $A$ is
a simple-current and $Ao=(o',o'')$.

This argument also showed that the quantum-dimension $S_{(\la,\mu)o}/S_{0,o}$
equals the product of the WZW $\widehat{{\rm su}}(N)$ quantum-dimensions of
$\la$ and $\mu$ at levels $p-N$ and $p'-N$, respectively.
Since the only simple-currents of $\widehat{{\rm su}}(N)_k$ are the ones $J^i$ corresponding
to the order-$N$ cyclic symmetry of the $\widehat{{\rm su}}(N)$ Dynkin diagram [53],
the only simple-currents of the $W_N$ minimal models are the $(J,J)$-orbit
of those primaries $(J^i,id)$. Hence $A=(J^i,id)$ for some $i$.

In fact, we know the order of $A=J_o$ must be 1 or 2 (see e.g.\ Consequence 2(i),
although this can be easily seen directly). Thus when $N$ is odd, $J_o=id$,
and when $N$ is even, the only possibilities are $J_o=id$ or $J_o=(J^{N/2},id)$.

\smallskip Next, let's give a proof of {Fact 3}.
By Consequence 2(ii) it is now automatic for $N$ odd, since then $J_o=id$.
So let $N$ be even and      $J_o=(J^{N/2},id)$.
Let $M$ be any modular invariant of $W_N$. From $MT=TM$ we see that
if $M_{(\la,\mu)(\la',\mu')}\ne 0$, then (16a) requires
$|p\la-p'\mu|^2 \equiv |p\la'-p'\mu'|^2$ (mod $2pp'$).
But an easy calculation shows (for su$(N)$) that
$N|\la|^2\equiv -t(\la)^2$ (mod $N$). Thus if $M_{(\la,\mu)(\la',\mu')}\ne 0$,
then the monodromies $Q_{J_o}=N Q/2$ of $(\la,\mu)$ and $(\la',\mu')$
are equal mod 1. Note then from $M=SMS^*$ that
$$\eqalign{M_{J_o,J_o}=&\,\sum_{(\la,\mu),(\la',\mu')} S_{J_o,(\la,\mu)}\,M_{(\la,\mu)
(\la',\mu')}\,S_{(\la',\la'),J_o}^*\cr =&\,
\sum_{(\la,\mu),(\la',\mu')} S_{0,(\la,\mu)}\,M_{(\la,\mu)
(\la',\mu')}\,S_{(\la',\la'),0}^*=M_{00}=1\ .}$$
Thus by (6c) and the fact that $o=\sigma_\ell J_o$, we get that $M_{oo}=1$,
which is {Fact 3}.

\smallskip We conclude this section, and this paper, with the proof of
Consequence 4. It is helpful to remember that the simple-current $J_o$ for
$W_3$ must be trivial, so all we have to account for here is the Galois
automorphism. The minimal $(p,4)$ $W_3$ models have unitarisation $\widehat{S}$
equal to the modular $S$ matrix of
the affine algebra $\widehat{{\rm su}}(3)$ at level $p-3$. Explicitly,
the primary $\la=(\la_0,\la_1,\la_2)\in P^{p-3}_{+}(\widehat{{\rm su}}(3))$ corresponds to the
$W_3$ primary given by the $(J,J)$-orbit $[\lambda+\rho]:=(J^i(2,1,1),J^i(\la_0+1,\la_1+1,\la_2+1))$. The $T$ matrix enters into the modular invariant classification only
via the selection rule
$$M_{ab}\ne 0\ {\rm iff}\ h_a\equiv h_b\ ({\rm mod}\ 1)\ ,\ \forall a,b\in\Phi\ ,$$
and in both our cases $T$-invariance reduces to (in the $W_3$ notation) the
selection rule
 $$M_{[\lambda][\mu]}\ne 0\ {\rm iff}\ \la_1^2+\la_1\la_2+\la_2^2\equiv 
\mu_1^2+\mu_1\mu_2+\mu_2^2\ ({\rm mod}\ 3p)\ ,\ \forall \la,\mu\in P_{++}^p\ .$$
Thus, using that primary field correspondence, the modular invariants and {\smcap nim}-reps of
these two different pairs of modular data can be identified. The modular
invariants are classified in [54]. Of course the exceptional $\widehat{{\rm su}}
(3)$ modular invariants all occur for odd level and so don't arise here.
Many {\smcap nim}-reps for $\widehat{{\rm su}}(3)$ were first given in [55], and other
{\smcap nim}-reps were obtained in other papers, culminating with the announcement
in [56] that there is a unique {\smcap nim}-rep for each $\widehat{{\rm su}}(3)$ modular
invariant. Ocneanu's argument requires the full structure of his von Neumann
subfactor theory (see e.g.\ [57] for a review) and it is not yet known if all 
this has a {\it necessary} counterpart in
RCFT. In any case a {\smcap nim}-rep for each $\widehat{{\rm su}}(3)$ modular invariant
(and hence each $W_3$ $(p,4)$ model) can be found in e.g.\ [40]. 

More generally, the $W_N$ minimal models at $(p,p')=(p,N+1)$
will for most $p$ correspond to the WZW $\widehat{{\rm su}}(N)$ model at level
$p-N$.

\bigskip\noindent{{\bf 7. Concluding remarks}}\medskip

This paper compares nonunitary and unitary RCFT. In a nonunitary theory, some 
of the  properties we would like to ascribe to the vacuum 0 instead belong
to what we call the {\it minimal primary} $o$, which corresponds to the
positive column of the modular $S$ matrix. 

\medskip\noindent{{\bf Question 1.}} Must there be exactly one primary with
minimal conformal weight?\medskip

In all examples of (healthy) nonunitary RCFTs known to this author, the answer
to Question 1 is `yes' and it is tempting to conjecture that it always must be.
 The affine algebras at fractional level can have 
several different admissible highest weight representations with minimal 
conformal weight (see e.g.\ [49]), but they don't have a direct interpretation as an RCFT
and so don't constitute a counterexample. When there is only one primary with
minimal conformal weight, that primary must be the minimal primary $o$ (hence the name).

A fundamental question here is
the multiplicity of the minimal primary $o$ in the full RCFT. We showed that this
multiplicity must be 1 for any minimal $W_N$ model, and we conjectured in \S4
 that it always equals 1.

The relation between 0 and $o$  lies at the heart of this paper.
We identify a property which many (but not all) RCFTs obey, namely that
the minimal primary $o$ and the vacuum 0 in the chiral theory are related to 
each other by what we call the Galois shuffle. Once again, all minimal $W_N$
models obey this {\it GS property}.  

\medskip\noindent{{\bf Question 2.}} Is there a characterisation of the
nonunitary RCFTs which obey the GS property? How typical is it?\medskip

Our reasons for suspecting that it's quite typical are that we know many
RCFTs which obey it, and also that many W-algebras can be constructed from
the affine algebras at fractional level, and the latter respects it (see
\S5). On the other
hand Consequence 2(iv) in \S4 indicates that it is quite nontrivial.

We learned in \S4 that the GS property has many nice consequences.
Roughly speaking, it means that 
the nonunitary theory behaves almost like a unitary theory. That corresponding
theory (or rather its modular data) is called the `unitarisation' of the 
unitary theory. The unitarisation obeys the same fusion rules as the 
nonunitary theory, and their toroidal
and cylindrical partition functions will be in one-to-one correspondence. 

\medskip\noindent{{\bf Question 3.}} Can we see the unitarisation {\it directly} inside
the nonunitary RCFT? In particular, it makes sense at the chiral level, so 
can we see it at the level of the chiral algebras? \medskip

\noindent{{\bf Question 4.}} Will the unitarisation always equal the modular data of
 a completely healthy unitary RCFT?\medskip

As mentioned in \S4, the unitarisation will always satisfy conditions (3)
and the identities in [14], and all the properties of unitary modular data, but 
nevertheless it still might not be realised as the modular data of an actual theory.
 A simple critical case is given
by the Virasoro minimal models at $(p,2)$, where $p\ge 1$ is odd: we find
in \S6 that its unitarisation can be expressed as
$$\eqalign{\widehat{S}_{ab}=&\,{2\over\sqrt{p}}\sin(\pi {ab\over p})\ ,\cr
\widehat{T}_{ab}=&\,\delta_{ab}\exp[-\pi\,({\pm a^2\over 2p}+{3\pm 1 \over 12})]\ ,}$$
where the primaries $a,b$ consist of all odd numbers $1\le a,b\le p-2$, and
where we take the upper signs (i.e.\ `+') in the formula for $T$ if
$p\equiv +1$ (mod 4), and otherwise take the lower signs (i.e.\ `$-$').
The question is, can this modular data be realised by the characters of
an RCFT. It obeys all properties (e.g.\ Fact 1) this author has been able
to check. For $p=3,5,9,11$, this data coincides with that of the WZW models
with affine algebras $\widehat{E_8\oplus E_8\oplus E_8}$ at level (1,1,1), $\widehat{F}_4$
at level 1, $\widehat{G}_2$ at level 2, and $\widehat{F}_4$ at level 2.
For the other values of $p$, this author has been unable to find a healthy
RCFT which realises it (but that certainly doesn't mean that none exists).
Curiously, for arbitrary (odd) $p$, that matrix $S$ coincides with the matrix $\psi$
diagonalising (in the sense of eq.(4) above) the spurious $\widehat{{\rm su}}(2)$ level $p-2$ {\smcap nim}-rep
called the tadpole in [55]. In any case, if an RCFT realisation can be found for
this modular data for all odd $p$, this would lend support to the thought that
the uniformisation of a nonunitary RCFT is the modular $S$ and $T$ matrices
for a healthy unitary RCFT.

Another reason the existence of a unitarisation  might be interesting is the 
following.
A deep relationship between von Neumann subfactors and RCFT has been developed
by Ocneanu, Evans, and others (see e.g.\ [57,36] for reviews).
Although subfactors cannot recover the entire RCFT (e.g.\ they don't
see the chiral algebra or even its character), they
 can realise for example the fusions, the modular $S$- and
$T$-matrices, the 1-loop modular invariant partition functions, and the
{\smcap nim}-reps of RCFTs. The subfactor picture (at present) 
can only realise {\it unitary} RCFTs. It is tempting to speculate 
that perhaps  subfactors realise instead the unitarisation $\widehat{S},\widehat{T}$
and corresponding modular invariants and {\smcap nim}-reps, even if
the unitarisation were not to correspond to a completely healthy RCFT.

\medskip\noindent{{\bf Question 5.}} Not all RCFTs have a unitarisation. However,
given any nonunitary RCFT, can we always find a unitary RCFT with identical
fusion rules?\medskip

Again, this is true of all examples known to this author. Of course it holds for
any RCFT obeying the GS property.

\bigskip\noindent{{\bf Acknowledgements}}\medskip

I thank David Evans, Matthias Gaberdiel, Christoph Schweigert,
Mark Walton and Jean-Bernard Zuber for  helpful comments.
This research was supported  by NSERC.

\bigskip
\noindent{{\bf References}}\medskip

\item{[1]} C.\ Schweigert, J.\ Fuchs, J.\ Walcher, ``Conformal field theory,
boundary conditions, and applications to string theory'', hep-th/0011109.

\item{[2]} W.\ Eholzer, M.\ R.\ Gaberdiel, Commun.\ Math.\ Phys.\ {\bf 186}
(1997) 61.

\item{[3]} D.\ Gepner,  E.\ Witten, Nucl.\ Phys.\ {\bf B278}
(1986) 493.

\item{[4]} J.\ Fuchs, C.\ Schweigert, {\it Symmetries, Lie Algebras
and Representations}, Cambridge Univ.\ Press, 1997.

\item{[5]} P.\ Goddard, A.\ Kent, D.\ Olive, Commun.\ Math.\ Phys.\
{\bf 103} (1986) 105.

\item{[6]} V.G.\ Kac, M.\ Wakimoto, Proc.\ Natl.\ Acad.\ Sci.\ USA
{\bf 85} (1988) 4956.

\item{[7]} A.\ Cappelli, C.\ Itzykson, J.-B.\ Zuber, Commun.\ Math.\
Phys.\ {\bf 13} (1987) 1.

\item{[8]} P.\ Bouwknegt, K.\ Schoutens, Phys.\ Rep.\ {\bf 223}
(1993) 183.

\item{[9]} G.\ Moore, N.\ Seiberg, Commun.\ Math.\ Phys.\ {\bf 123} (1989)
177.

\item{[10]} E.\ Verlinde, Nucl.\ Phys.\ {\bf B300} (1988) 360.

\item{[11]} J.\ Fuchs, Fortsch.\ Phys.\ {\bf 42} (1994) 1.

\item{[12]} T.\ Gannon, ``Modular data: the algebraic combinatorics of
conformal field theory'', math.QA/0103044.

\item{[13]} P.\ B\'antay, Phys.\ Lett.\ {\bf B394} (1997) 87.

\item{[14]} P.\ B\'antay, ``On generalizations of Verlinde's formula'',
hep-th/0007164.

\item{[15]} J.\ Cardy, Nucl.\ Phys.\ {\bf B270} (1986) 186.

\item{[16]} J.\ Cardy, Nucl.\ Phys.\ {\bf B324} (1989) 581.

\item{[17]} G.\ Pradisi, A.\ Sagnotti, Y.\ S.\ Stanev, Phys.\
Lett.\ {\bf B381} (1996) 97.

\item{[18]} R.\ E.\ Behrend, P.\ A.\ Pearce, V.\ B.\ Petkova,
J.-B.\ Zuber, Phys.\ Lett.\ {\bf B444} (1998) 163.

\item{[19]} J.\ Fuchs, C.\ Schweigert, Nucl.\ Phys.\ {\bf B530}
(1998)  99.

\item{[20]} A.\ Recknagel, V.\ Schomerus, Nucl.\ Phys.\  {\bf B531}
(1998) 185.

\item{[21]} {A.\ Coste, T.\ Gannon},  Phys.\ Lett.\ {\bf B323} (1994) 316.
 
\item{[22]} P.\ B\'antay, Commun.\ Math.\ Phys.\ {\bf 233} (2003) 423.

\item{[23]} A.\ Coste, T.\ Gannon, ``Congruence subgroups and RCFT'',
math.QA/9909080.

\item{[24]} T.\ Gannon, Nucl.\ Phys.\ {\bf B627} (2002) 506.

\item{[25]} D.\ Gepner, ``Foundations of rational quantum field
theory, I'', hep-th/9211100.

\item{[26]} A.\ B.\ Zamolodchikov, Theor.\ Math.\ Phys.\ {\bf 65}
(1986) 1205.

\item{[27]} P.\ Goddard, In: {\it Infinite dimensional Lie algebras and
groups,} ed.\ by V.G.\ Kac, (World Scientific, 1989) 237.

\item{[28]} A.\ Medina, P.\ Revoy, Ann.\ scient.\ \'Ec.\ Norm.\ Sup.\ {\bf 18}
(1985) 553. 

\item{[29]} J.M.\ Figueroa-O'Farrill, S.\ Stanciu, J.\ Math.\ Phys.\ {\bf 37}
(1996) 4121.

\item{[30]} E.\ Witten, Nucl.\ Phys.\ {\bf B311} (1989) 46.

\item{[31]} C.R.\ Nappi, E.\ Witten, Phys.\ Rev.\ Lett.\ {\bf 71} (1993) 3751.

\item{[32]} B.\ H.\ Lian, Commun.\  Math.\ Phys.\ {\bf 163} (1994) 307.

\item{[33]} A.\ Giveon, O.\ Pelc, E.\ Rabinovici, Nucl.\ Phys.\ {\bf B462}
(1996) 53.

\item{[34]} F.\ R.\ Gantmacher, {\it The Theory of Matrices}, Chelsea
Publishing, New York, 1990.

\item{[35]} A.N.\ Schellekens, S.\ Yankielowicz, Phys.\ Lett.\ {\bf
B227} (1989) 387.

\item{[36]} J.\ B\"ockenhauer, D.E.\ Evans, ``Modular invariants from
subfactors'', math.OA/0006114.

\item{[37]} S.\ Lu, Phys.\ Lett.\ {\bf B218} (1989) 46.

\item{[38]} E.\ Lapalme, Y.\ Saint-Aubin, J.\ Phys.\ {\bf A34} (2001) 1825.

\item{[39]} P.\ Ruelle, J.\ Phys.\ {\bf A32} (1999) 8831.

\item{[40]} R.\ E.\ Behrend, P.\ A.\ Pearce, V.\ B.\ Petkova,
J.-B.\ Zuber, Nucl.\ Phys.\ {\bf B579} (2000) 707.

\item{[41]} E.\ Beltaos, T.\ Gannon, work in preparation.

\item{[42]} A.Ch.\ Ganchev, V.B.\ Petkova, G.\ Watts, Nucl.\ Phys.\
{\bf B571} (2000) 457.

\item{[43]} P.\ Mathieu, M.\ A.\ Walton, Prog.\ Theor.\ Phys.\ Suppl.\
{\bf 102} (1990) 229.

\item{[44]} H.\ Awata, Y.\ Yamada, Mod.\ Phys.\ Lett.\ {\bf A7} (1992)
1185.

\item{[45]} B.\ Feigin, F.\ Malikov,  Lett.\ Math.\ Phys.\ {\bf 31} (1994) 315.

\item{[46]} M.\ R.\ Gaberdiel, Nucl.\ Phys.\ {\bf B618} (2001) 407.

\item{[47]}  F. Lesage, P. Mathieu, J. Rasmussen, H. Saleur,  Nucl.\ 
Phys.\ {\bf B647} (2002) 363.

\item{[48]} S.\ Ramgoolam, ``New modular Hopf algebras related to
rational $k$ $\widehat{sl(2)}$'', hep-th/9301121.

\item{[49]} P.\ Mathieu, D.\ S\'en\'echal, M.A.\ Walton, Int.\ J.\
Mod.\ Phys.\ {\bf A7} Suppl.\ 1B (1992) 731.

\item{[50]} A.\ Rocha-Caridi, In: {\it Vertex Operators in Mathematics and
Physics}, ed. by J.\ Lepowsky, S.\ Mandelstam and I.M.\ Singer, (Springer, 1985) p.451.

\item{[51]} V.A.\ Fateev, S.\ Lykyanov, Int.\ J.\ Mod.\ Phys.\ {\bf
A3} (1988) 507.

\item{[52]} E.\ Frenkel, V.\ Kac, M.\ Wakimoto, Commun.\ Math.\ Phys.\
{\bf 147} (1992) 295.

\item{[53]} J.\ Fuchs, Commun.\ Math.\ Phys.\ {\bf 136} (1991) 345.

\item{[54]} T.\ Gannon, Commun.\ Math.\ Phys.\ {\bf 161} (1994) 233.

\item{[55]} P.\ Di Francesco, J.-B.\ Zuber, Nucl.\ Phys.\ {\bf B338} (1990)
602.

\item{[56]} A.\ Ocneanu, talk, Kyoto 2000.

\item{[57]} D.\ E.\ Evans, Y.\ Kawahigashi, {\it Quantum Symmetries on
Operator Algebras}, Oxford University Press, 1998.

\end